\documentclass[twocolumn,times,tighten]{aastex62}
\usepackage{amsmath,amstext}
\usepackage[english]{babel}
\usepackage[utf8x]{inputenc}

\newcommand{\myedit}[1]{{#1}}

\newcommand{\teff}{$T_\mathrm{eff}$}
\newcommand{\logg}{$\log g$}
\newcommand{\feh}{[Fe/H]}

\shorttitle{Chemistry of the Galactic Center}
\shortauthors{Thorsbro et al.}

\begin{document}



\title{Evidence against anomalous compositions for giants in the Galactic Nuclear Star Cluster}


\author[0000-0002-5633-4400]{B. Thorsbro}
\affil{Lund Observatory, Department of Astronomy and Theoretical Physics, Lund University, Box 43, SE-22100 Lund, Sweden}
\email{thorsbro@astro.lu.se}

\author[0000-0001-6294-3790]{N. Ryde}
\affil{Lund Observatory, Department of Astronomy and Theoretical Physics, Lund University, Box 43, SE-22100 Lund, Sweden}

\author[0000-0002-6590-1657]{M. Schultheis}
\affil{Observatoire de la C\^ote d'Azur, CNRS UMR 7293, BP4229, Laboratoire Lagrange, F-06304 Nice Cedex 4, France}

\author[0000-0001-9853-2555]{H. Hartman}
\affil{Malm\"o University, Materials Science and Applied Mathematics, SE-20506 Malmö, Sweden}
\affil{Lund Observatory, Department of Astronomy and Theoretical Physics, Lund University, Box 43, SE-22100 Lund, Sweden}

\author[0000-0003-0427-8387]{R. M. Rich}
\affil{Department of Physics and Astronomy, UCLA, 430 Portola Plaza, Box 951547, Los Angeles, CA 90095-1547, USA}

\author[0000-0001-8161-1732]{M. Lomaeva}
\affil{Lund Observatory, Department of Astronomy and Theoretical Physics, Lund University, Box 43, SE-22100 Lund, Sweden}

\author[0000-0002-6040-5849]{L. Origlia}
\affil{INAF---Osservatorio Astronomico di Bologna, via Gobetti 93/3, I--40129 Bologna, Italy}

\author[0000-0002-4912-8609]{H. J\"onsson}
\affil{Lund Observatory, Department of Astronomy and Theoretical Physics, Lund University, Box 43, SE-22100 Lund, Sweden}

\begin{abstract}
Very strong \ion{Sc}{1} lines have been found recently in cool M~giants in the Nuclear Star Cluster in the Galactic Center. Interpreting these as anomalously high scandium abundances in the Galactic Center would imply a unique enhancement signature and chemical evolution history for nuclear star clusters, and a potential test for models of chemical enrichment in these objects. We present high resolution K-band spectra (NIRSPEC/Keck~II) of cool M~giants situated in the solar neighborhood and compare them with spectra of M~giants in the Nuclear Star Cluster. We clearly identify strong \ion{Sc}{1} lines in our solar neighborhood sample as well as in the Nuclear Star Cluster sample. The strong \ion{Sc}{1} lines in M~giants are therefore not unique to stars in the Nuclear Star Cluster and we argue that the strong lines are a property of the line formation process that currently escapes accurate theoretical modeling. We further conclude that for giant stars with effective temperatures below approximately $3800$\,K these \ion{Sc}{1} lines should not be used for deriving the scandium abundances in any astrophysical environment until we better understand how these lines are formed. We also discuss the lines of vanadium, titanium, and yttrium identified in the spectra, which demonstrate a similar striking increase in strength below $3500$\,K effective temperature.
\end{abstract}

\keywords{stars: abundances --- late-type --- Galaxy: center}

\section{Introduction}

With the advent of high resolution infrared spectroscopy, it has become possible to explore the spectra and composition of stars in the nuclear star cluster (NSC) just a few parsecs from the Galactic Center. Several chemical abundance studies have addressed the giants in the Galactic Center region  and the NSC \citep[see e.g.,][]{ryde_schultheis:15,rich:17,do:18}. Spectroscopy in the Galactic Center poses a special challenge, as the high extinction generally at present restricts investigations to the infrared K band.  Although there is considerable heritage in the 1.6 $\mu \rm m$ H band from e.g. APOGEE and earlier studies using NIRSPEC/Keck~II \citep[e.g.][]{origlia:11}, one is challenged by the paucity of weak lines suitable for abundance analysis, as well as the presence of molecular bands that cause blends.  The cool, luminous, giants of the NSC are easiest to observe, but pose the greatest perils for analysis.

Low resolution studies have advanced a scenario in which the NSC and nuclear disk have abundant metal rich stars, reaching to [Fe/H] $=+1$ \citep{do:15,Feldmeier-Krause2017}. \citet{rich:17} challenges this picture with new high resolution NIRSPEC spectra in the Galactic Center, and finding no stars above [Fe/H] $=+0.6$. \citet{do:18} reports high resolution spectra of NSC stars behind AO correction, arguing for extreme enhancements of scandium, vanadium, and yttrium; in some cases the analysis finds 10 times Solar abundance for these elements.


Especially interesting are the strong \ion{Sc}{1} lines found in M~giants that are discussed in \citet{rich:17}. They suggest non-LTE effects as the cause for them, while \citet{do:18} argue for extreme over-abundance (as much as a factor of ten compared to iron) of scandium in the NSC. If confirmed this latter interpretation would be a chemical signature of the special environment in the Galactic Center, and potentially very important. There are good reasons to assume that the enrichment and star formation histories might be different in the Galactic Center, especially if the NSC has a unique formation history. Large magnetic fields, suppressed star formation, high turbulence, tidal forces, and the deep Galactic Center potential well might lead to a very different chemical history for the stellar populations in the Galactic Center.  Furthermore, one might expect inhomogeneities in the trends with a larger scatter including outliers due to a possible mixture of populations that in principle might include the disk, inner, halo, NSC, nuclear disk, and bulge.  A unique scandium abundance trend for the NSC would suggest that the Galactic Center and similar environments is a site for a new channel of nucleosynthesis of scandium and possibly other elements. Such a trend might also provide a powerful chemical tag for stars formed in nuclear environments.

Scandium resides between the $\alpha$-elements calcium and titanium in the periodic table and is considered an iron-group element. Even titanium is sometimes considered an iron-group element, such as in the discussion of the metal-poor star HD84937 \citep{sneden:16}; therefore scandium can be seen as an intermediate element between the $\alpha$ elements and the iron-peak group. The precise origin of scandium and its only stable isotope, $^{45}$Sc, seems to be complex and is still a matter of debate.  Scandium is produced in the innermost ejected layers of core-collapse SNe (type II) during neon burning and explosive silicon and oxygen burning via the radioactive progenitor $^{45}$Ti, as reviewed by \citet{woosley:95,romano:10}, while the contribution from type Ia SNe seems to be negligible \citep{iwamoto:1999,clayton:03}. However, the predicted trends of scandium with [Fe/H] disagree even when taking into account metallicity- and mass-dependent yields, which might be important \citep{woosley:95,nomoto:13,chieffi:02}. Chemical evolution models predict for too little scandium production. This could be due to the problems in the stellar yield calculations (see also \citealt{romano:10}). It should be noted, however, that although extreme enhancements of individual elements are known in stars of low metallicity, for stars with [Fe/H]$>$0, the total metal production is so high that no single supernova event can affect the abundance of a given species, save perhaps for an r-process production event that might result from a neutron star merger. These factors raise the bar significantly for any purported enhancements of metals in stars of high metallicity.

From the observational point of view scandium seems to behaves like a typical $\alpha$-element, e.g.\ enhanced in the thick disk and the galactic Bulge \citep{Battistini15,lomaeva:18}. \citet{nissen:2000} and \citet{howes:2016} also find [Sc/Fe]$\approx +0.3$ abundances for halo stars and metal poor stars toward the bulge, however their studies have no stars more metal rich than [Fe/H]$\sim-1.5$.  It is noteworthy that \citet{gratton:1991,prochaska:2000,ernandes:2018} report a constant $\rm [Sc/Fe] \sim 0$ (and as well for V) for galactic Bulge globular clusters spanning $-1.5 < \rm {[Fe/H]} < 0.0$. \myedit{\cite{smith:2002} report a solar mean scandium abundance for 12 red giants all having subsolar metallicity in the Large Magellanic Cloud---an investigation carried out in the K-band.}





In the Galactic center, the picture has been more complicated.
\citet{carr:00} measured scandium abundance for the cool supergiant star IRS7 located in the Galactic Center and found extremely strong \ion{Sc}{1} lines as well as \ion{V}{1} and \ion{Ti}{1} lines. An abundance of $\rm [Sc/Fe] \sim 0.9\,dex$ is required to fit the strength of the \ion{Sc}{1} lines. However, supergiant stars are affected by large velocity fields, depth-dependent turbulence, temperature inhomogeneities, etc., and it is known that fully realistic atmospheres for supergiants remain a challenge.

We have employed high resolution K band spectroscopy to overcome the high and variable extinction toward the NSC \citep[see, e.g.,][]{ryde:16b,rich:17}. \myedit{High-resolution, K-band spectroscopy was pioneered already in the late 80s by Smith \& Lambert \citep{smith:85,smith:86,smith:90}.} We can now push fainter to observe cool M~giants with the largest telescopes, and thus avoid the supergiant stars. However, as we emphasized earlier, even the interpretation of the M~giant spectra is challenging. Our aim here is to test whether the strong scandium lines in cool M~giants in the NSC are due to either physical effects in the line-formation process \citep{rich:17} or due to intrinsically high scandium abundances \citep{do:18}. Toward this aim, we have acquired spectra for range of solar neighborhood stars, similar to those we have observed in the NSC in \citet{rich:17}, using the same instrument and telescope configuration (NIRSPEC on KECK II). These are used as a benchmark to compare with and we discuss different possible reasons for the strong \ion{Sc}{1} lines in the K-band spectra of M giants.

\section{Observations}

The high resolution, K-band spectra have been obtained using the NIRSPEC \citep{nirspec_mclean,mclean} facility at Keck~II, using the 0.432"x 12" slit and the NIRSPEC-7 filter, giving the resolving power of $R\sim23,000$ needed for accurate abundance determination. 5 spectral orders are recorded, covering the wavelength range of 21,000-24,000\,\AA. However, the wavelength coverage is not complete; there are gaps between the orders.

Apart from the 18 stars observed in the Galactic Center, which are presented in \citet{ryde:16,rich:17}, seven M giants in the solar neighborhood vicinity have been observed using nirspec on Keck II using the configuration described above, on 28-29 July 2017, under program U103NS (PI: Rich). These stars are selected to be of the same spectral type as the stars observed in the Nuclear Star Cluster. Table~\ref{tab:star} provides the basic data for the new observations including the apparent K$_\text{s}$ band magnitudes. For details about the data reduction process we refer to \citet{rich:17}.

\begin{deluxetable*}{l c r r r c r c c c }[t]
\tablecaption{Compiled data for both the observed stars and stars used for comparisons in this paper. We assume solar abundances of A(Fe) = 7.45 \citep{solar:sme} and A(Sc) = 3.04 \citep{pehlivan:sc17}. \label{tab:star}}
\tablewidth{0pt}
\tablehead{
\colhead{Star} & \colhead{Obs.\ date} & \colhead{K$_\text{s}$} & \colhead{RA} &  \colhead{dec} & \colhead{\teff} & \colhead{\logg} &  \colhead{\feh} & \colhead{$\xi_\mathrm{micro}$} & \colhead{[Sc/Fe]} \\
   & & & \colhead{[h:m:s]} & \colhead{[d:m:s]} & [K] & & & [km\,s$^{-1}$] &
  } 
\startdata
2M17584888--2351011 & 2017 Jul 30 & \,~\,~6.49 & 17:58:48.89 & $-$23:51:01.17    & 3652 & 1.44 & \,~\,~0.27 & 2.0 & 0.80 \\
2M18103303--1626220 & 2017 Jul 29 & \,~\,~6.51 & 18:10:33.04 & $-$16:26:22.06    & 3436 & 0.79 & \,~\,~0.27 & 2.2 & 0.77 \\
2M18191551--1726223 & 2017 Jul 29 & \,~\,~6.56 & 18:19:15.51 & $-$17:26:22.35    & 3596 & 1.16 & \,~\,~0.21 & 2.0 & 0.54 \\
2M18550791+4754062  & 2017 Jul 30 & \,~\,~7.63 & 18:55:07.92 & \,~\,~47:54:06.22 & 3915 & 1.40 & $-0.27$ & 2.0 & 0.37 \\ 
2M19122965+2753128  & 2017 Jul 29 & \,~\,~6.60 & 19:12:29.66 & \,~\,~27:53:12.83 & 3263 & 0.25 & \,~\,~0.27 & 2.4 & 0.58 \\
2M19411424+4601483  & 2017 Jul 30 & \,~\,~7.69 & 19:41:14.25 & \,~\,~46:01:48.14 & 3935 & 1.41 & $-0.37$ & 2.0 & 0.43 \\ 
2M21533239+5804499  & 2017 Jul 29 & \,~\,~6.58 & 21:53:32.40 & \,~\,~58:04:49.94 & 3708 & 1.17 & \,~\,~0.29 & 2.0 & 0.16 \\
$\alpha$ Boo        & ---         &    $-$2.91 & 14:15:39.67 & \,~\,~19:10:56.67 & 4286 & 1.66 &    $-$0.52 & 1.7 & 0.16 \\
GC 7688             & ---         &   \,~11.00 & 17:45:42.17 &    $-$29:00:54.99 & 4150 & 1.78 &    $-$0.08 & 1.8 & 0.13 \\
GC 11025            & ---         &   \,~10.41 & 17:45:37.13 &    $-$29:00:14.39 & 3400 & 0.69 & \,~\,~0.27 & 2.3 & 1.27 \\
GC 11473            & ---         &   \,~11.74 & 17:45:42.64 &    $-$29:00:10.23 & 3550 & 1.13 & \,~\,~0.64 & 2.0 & 0.37 \\
\enddata
\end{deluxetable*}

\begin{figure*}[t]
\centering
\includegraphics[trim={0.5cm 0.8cm 0.2cm 4.8cm},clip,width=1.0\linewidth]{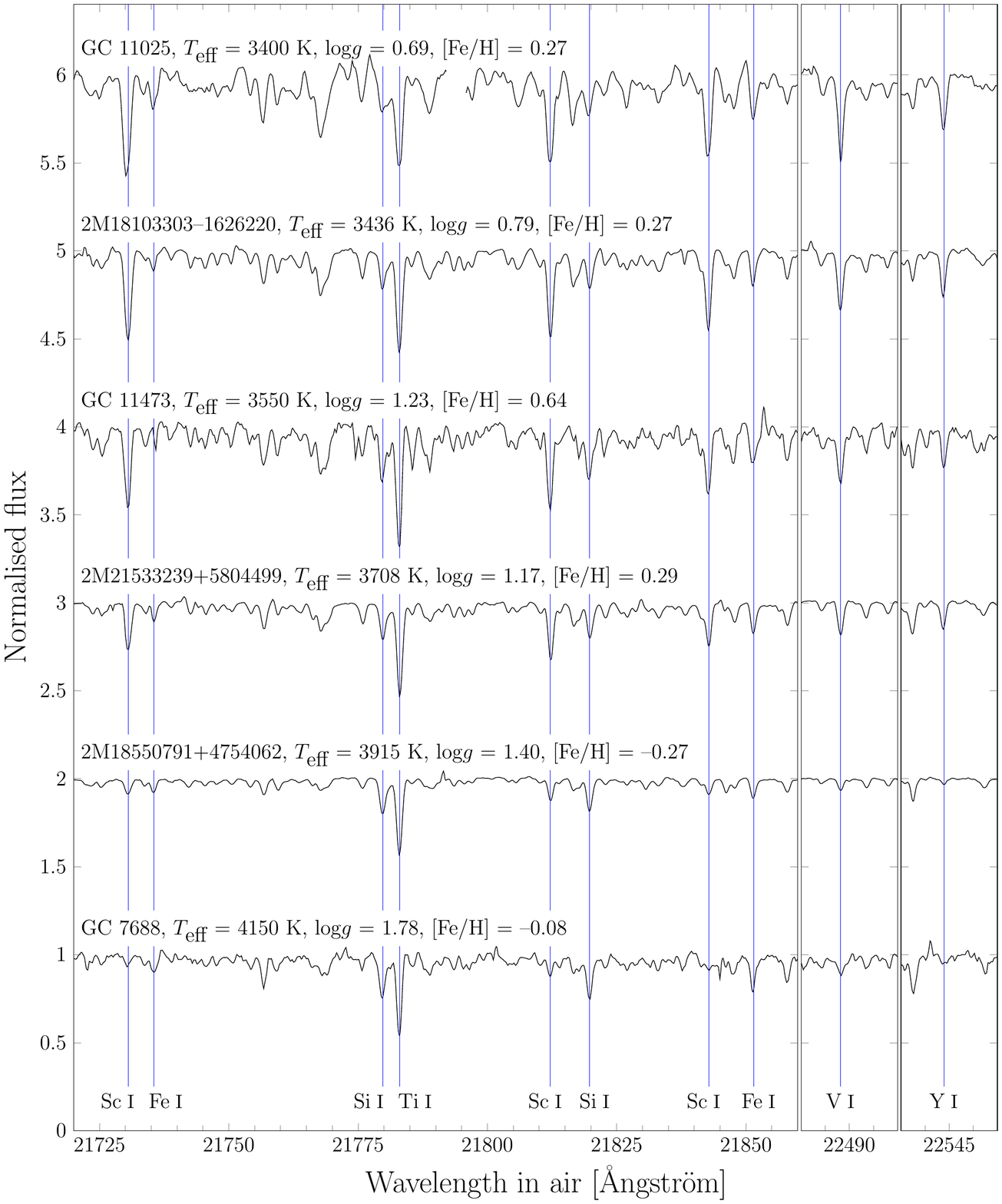}
\caption{Six M~giant stars plotted together with increasing temperature top to bottom with blue vertical lines identifying lines of interest. The normalized fluxes have been translated upwards with integer values for presentation. The stars are a mix of three Galactic Center stars (GC~7688, GC~11025, and GC~11473) and three solar neighborhood stars (2M18103303--1626220, 2M18550791+4754062, and 2M21533239+5804499). The spectra show striking strong scandium, vanadium, and yttrium lines in the cooler stars, even though the stars are located in widely different environments. As temperatures increase to $3900$\,K and beyond, the neutral lines of scandium, vanadium, and yttrium begin to vanish, presumably due to ionization.}
\label{2mvsgc}
\end{figure*}

Spectra of three observed solar neighborhood stars are plotted in Figure~\ref{2mvsgc} together with three Galactic Center stars from \citet{rich:17}. The stars have been ordered by temperature to illustrate the change in line strengths for scandium, vanadium, and yttrium which decrease dramatically from 3400 to 3800 K.  For scandium, we can illuminate this behavior by investigating the atomic physics in greater depth, something that we cannot yet do for the other features.

\section{Analysis}

\subsection{The Atomic Structure of \ion{Sc}{1}}\label{subsec:atomicsctruct}

In addition to the abundance of an \myedit{element}, there are other parameters affecting the line strengths of absorption lines. One is the oscillator strength and another is the population of the lower level. The latter depends strongly on the excitation energy. For most elements, the near-IR transitions appear at high excitation energies, making the level populations lower. This is the case for most iron-group elements such as \ion{Fe}{1}, where the lower excitation levels, such as the 4s levels, do not have near-IR transitions. For \ion{Sc}{1}, the structure is different, with low-excitation transitions 3d$^2$4s--3d4s4p appearing in the K-band. 

\begin{figure*}[t]
\centering
\includegraphics[height=88mm]{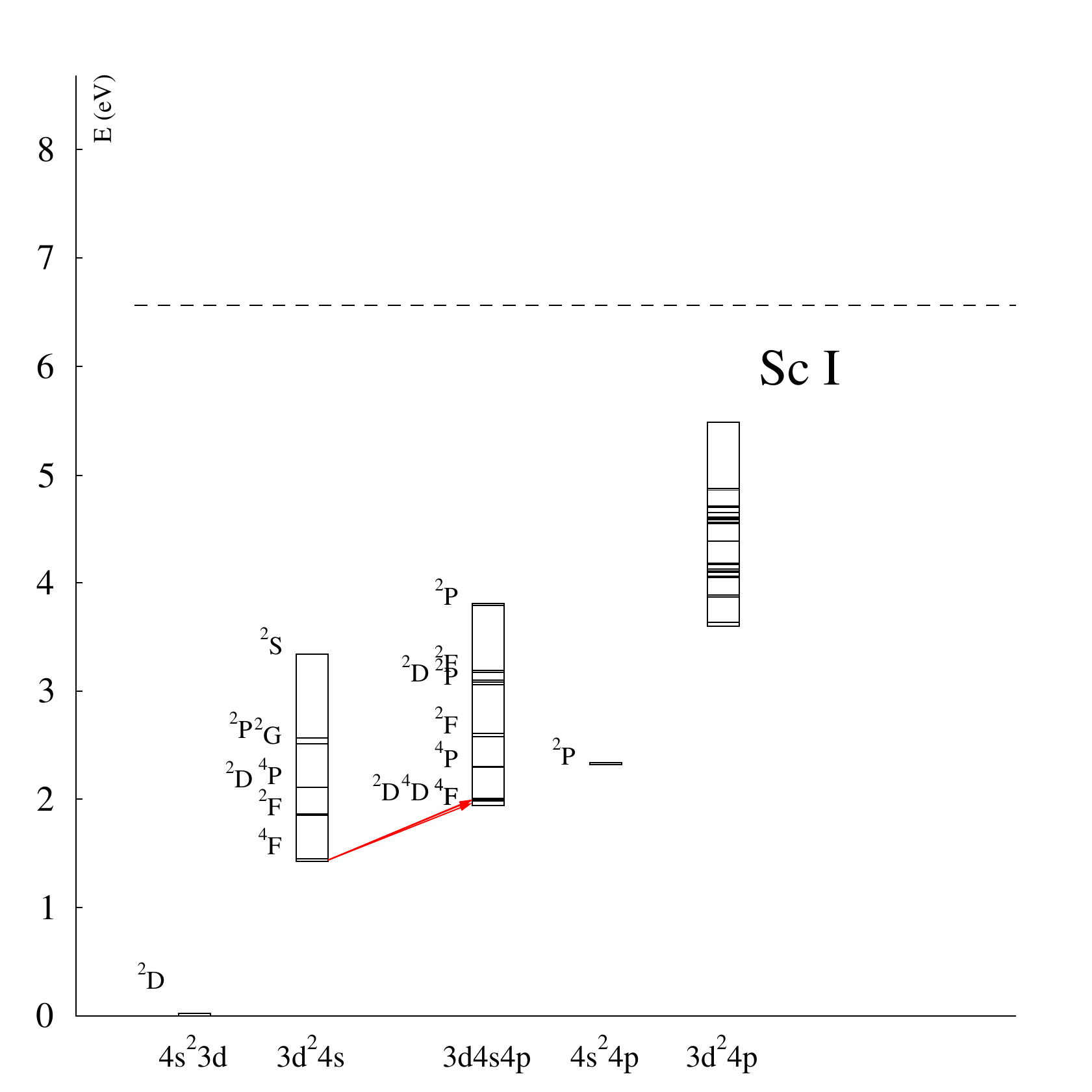}
\includegraphics[height=88mm]{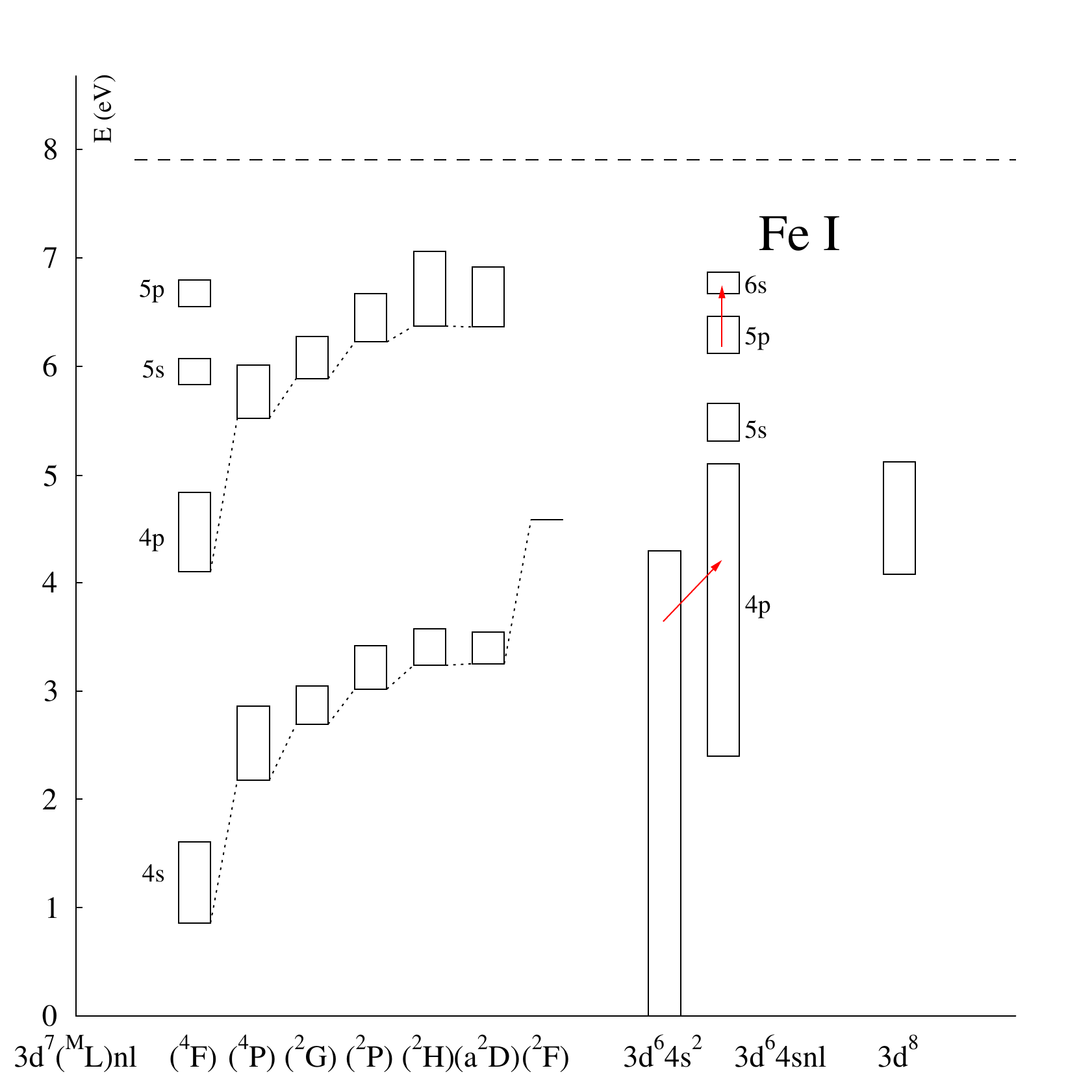}
\caption{The figure shows the energy level diagrams for \ion{Sc}{1} and \ion{Fe}{1}, respectively. The vertical energy scale is the same, and the dashed line shows the ionization energy. The observed lines are marked with red arrows. Note that the near-IR \ion{Sc}{1} lines are located at significantly lower excitation energies compared to the near-IR \ion{Fe}{1} lines.}
\label{scifei}
\end{figure*}

Figure \ref{scifei} shows the energy level diagrams for \ion{Sc}{1} and \ion{Fe}{1}, respectively, plotted with the same vertical energy scale. The dashed line shows the ionization energy for each atom. As can be seen, there is a significant difference in the excitation of the near-IR lines, where Sc I originates at much lower energies, around 1.5 eV, compared to Fe I at 3.6 and 6.1 eV. At temperatures of 3500-4500 K in the line-forming atmospheric depths, as for the stars in the current study, the Boltzmann factor, $e^{-\Delta E / \mathrm{k_b} T}$, gives a difference of 3 orders of magnitudes comparing the two different excitation energies and in this case the different atoms, which means a difference of 3 orders of magnitude for relative level population. Note, however, that the abundance difference between iron and scandium in a solar mixture of gas is more than 4 orders of magnitude. \myedit{A further implication is,
as noted early by \citet{smith:85,smith:90}, that non-LTE effects should be smaller for high-excitation lines, which are formed in deeper, warmer layers of the stellar atmosphere and lines from the dominant stage of ionization. Non-LTE effects could thus affect low-excitation lines from neutral species in M giants. It would, therefore, not be unexpected if non-LTE effects should plague the Sc abundances derived from our low-excitation Sc I lines.}

\subsection{Atomic Data of \ion{Sc}{1} Lines}

In general, there is a lack of experimental atomic data for near-IR transitions \citep{nailingthestars2013}. In response to this need, we have initiated a program to provide accurate and vetted near-IR atomic data for stellar spectroscopy. Scandium is one of the elements covered. Being an odd element, scandium has a non-zero nuclear spin of $I$=5/2 allowing for hyperfine structure (HFS). In recent works, we have measured the oscillator strengths and HFS of \ion{Sc}{1} \citep{pehlivan:sc,vandeelen:sc}. The oscillator strengths are derived using the radiative and lifetime method, and the uncertainties are 0.03~dex for the lines used in the present study. Since the HFS is a result of the interaction between the nuclear and electronic angular momenta, the effect is larger for electrons close to the nucleus. Unpaired s-electrons are thus expected to show the largest HFS.

The states responsible for the transitions used in the present study, 3d$^2$4s--3d4s4p, involve an unpaired 4s-electron, making the hyperfine structure of the lines used large. This is indeed what is observed in the laboratory measurements. The HFS-data used in the present study are derived from fitting Fourier Transform Spectroscopy (FTS) measurements \citep{vandeelen:sc}.

\subsection{Analyzing the effect of temperature and HFS}\label{teffeffect}

\begin{figure}[t]
\centering
\includegraphics[trim={0cm 0.6cm 0cm 2cm},clip,width=1.0\linewidth]{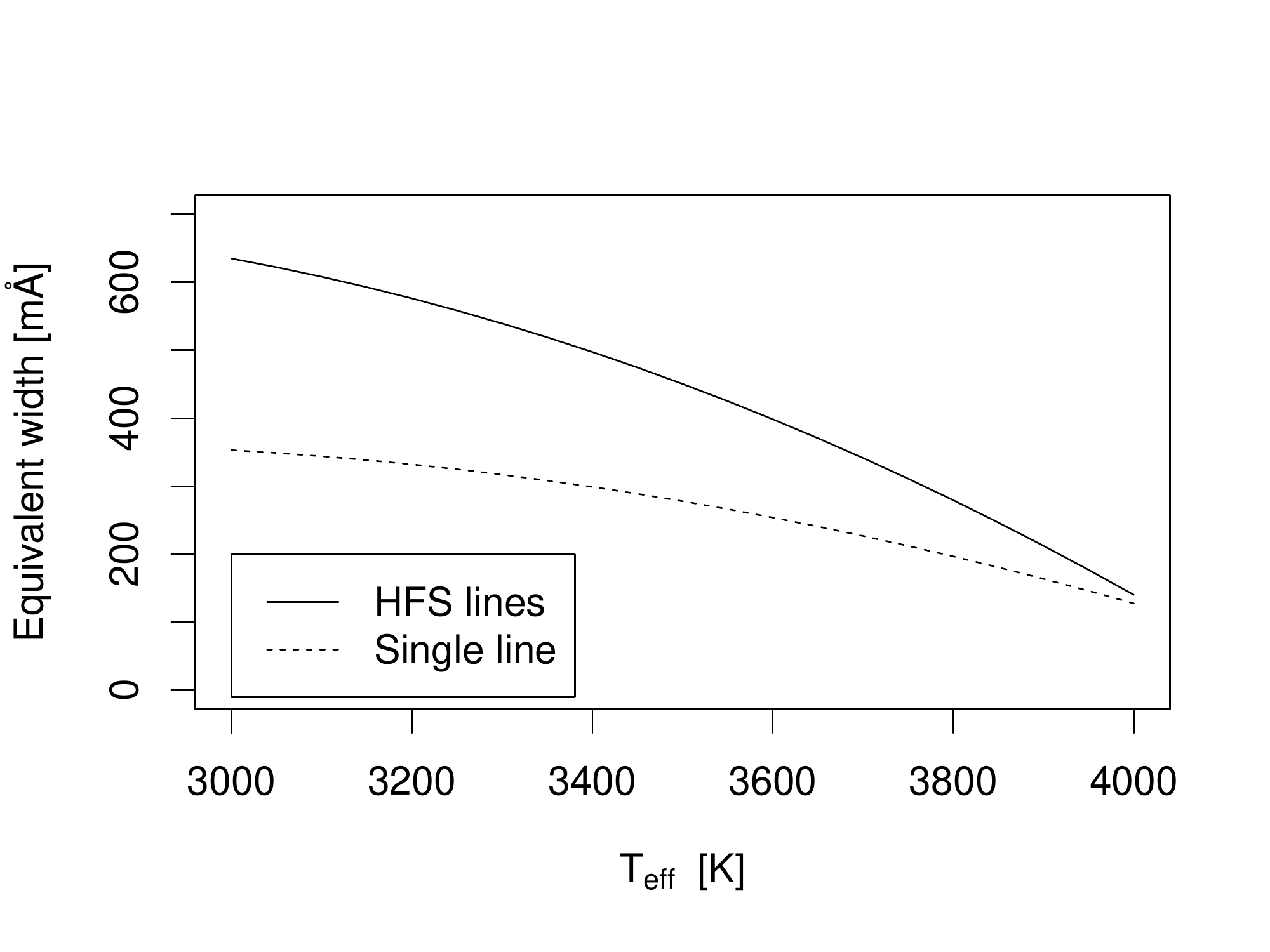}
\caption{The equivalent width of a \ion{Sc}{1} line as a function of temperature depends on whether the \ion{Sc}{1} line is considered to be a combination of many small lines due to hyperfine structure (HFS) or if the line is considered to be a singular atomic level transition. Notice, how at high temperature the two analyses converge, but at cooler temperatures the HFS of the line enables the spectral line feature to become considerably stronger compared to basing the analysis on a singular atomic level transition. The metallicity and scandium abundance is assumed to be solar, and the surface gravity is assumed to follow isochrone relations with changing temperatures. Non-LTE \myedit{and 3D} effects are not considered.}
\label{eqwidthvsteff}
\end{figure}

We model theoretical line formation of a spectral \ion{Sc}{1} lines to explore the effects of both temperature and hyperfine structure (HFS), using the BSYN \&  EQWIDTH codes based on routines from the MARCS model atmosphere code \citep{marcs:08}. We use one-dimensional (1D) MARCS atmosphere models which are hydrostatic model photospheres in spherical geometry, computed assuming LTE, chemical equilibrium, homogeneity, and conservation of the total flux (radiative plus convective, the convective flux being computed using the mixing-length recipe). The resulting line strength measured in equivalent width is plotted against temperature in Figure~\ref{eqwidthvsteff}. For the spectral line based on a single atomic level transition we use the measured oscillator strength from \citet{pehlivan:sc}. For the HFS based spectral line we use the combined work of \citet{pehlivan:sc,vandeelen:sc}. The fact that the HFS spectral line is a combination of many weak lines means that the spectral line does not saturate in the classical sense of a spectral line based on a single atomic transition, and thus can form a much stronger line.

\begin{figure}[t]
\centering
\includegraphics[trim={0cm 0.6cm 0cm 2cm},clip,width=1.0\linewidth]{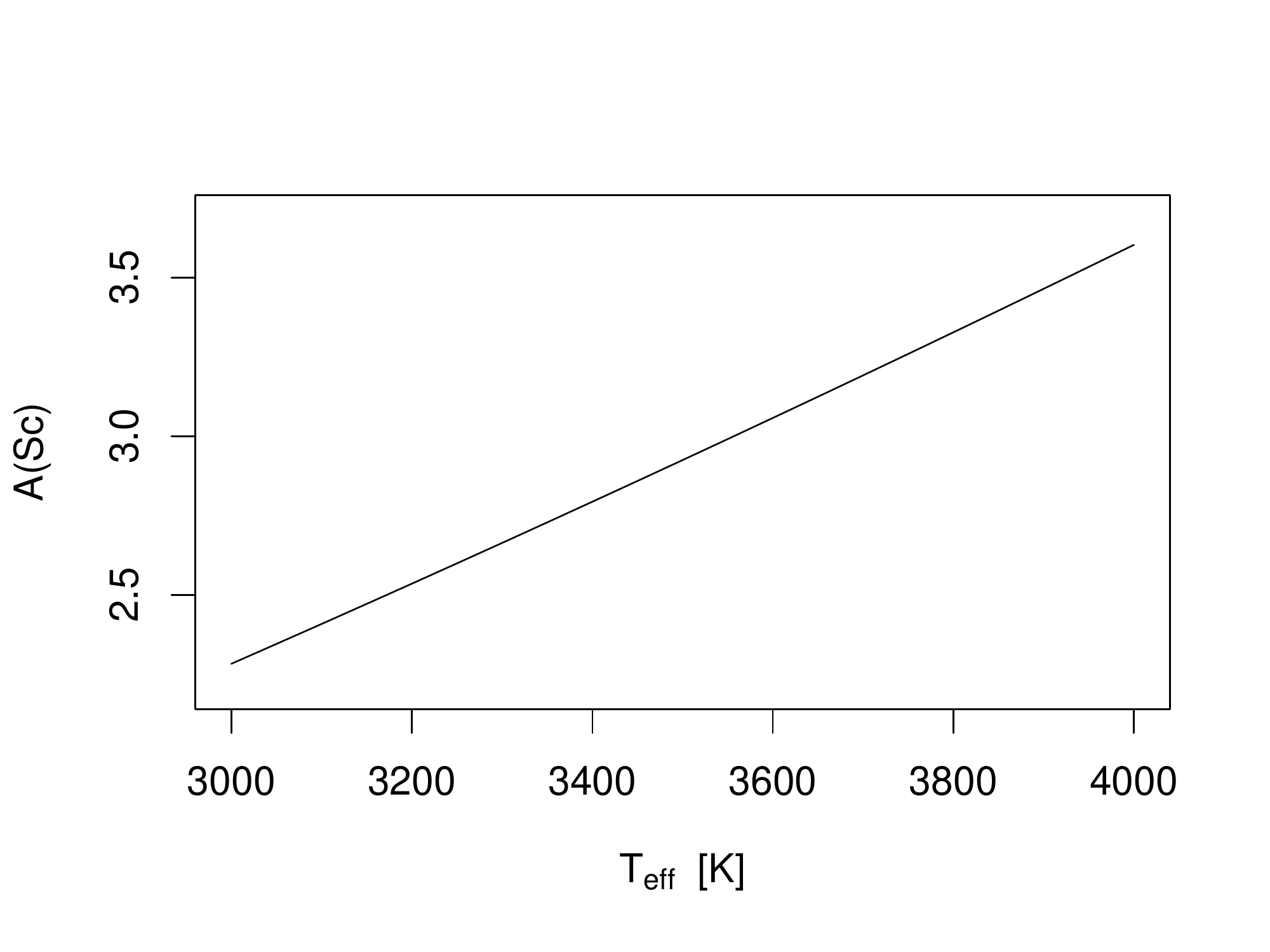}
\caption{This plot assumes a \ion{Sc}{1} line with a fixed equivalent width of 300~mÅ, as might be the situation when analyzing an observed spectrum. Plotted is the number density abundance of scandium as a function of temperature derived from this \ion{Sc}{1} line. The plot shows how cooler temperatures counteracts higher abundances when modeling such a feature. Note that a temperature difference of 100~K corresponds to an abundance difference of more than 0.1 dex. The metallicity is assumed to be solar, and the surface gravity is assumed to follow isochrone relations with varying temperature. Non-LTE \myedit{and 3D} effects are not considered.}
\label{scfevsteff}
\end{figure}

Using the same code and model assumptions, we further derive scandium abundances for a given spectral \ion{Sc}{1} line having an equivalent width of 300~mÅ, which is presented in Figure~\ref{scfevsteff}.

\subsection{Accurate Stellar Parameters}


The stellar parameters of the seven solar neighborhood stars presented here are obtained from the APOGEE Data Release 14 \citep{majewski:17,abolfathi:18,blanton:17}. The stellar parameters have been determined using the APOGEE stellar parameters and chemical abundance pipeline \citep[ASPCAP; ][]{garcia:16} and then calibrated using photometric effective temperatures, asteroseismic surface gravities, as well as stellar clusters \citep{holtzman:2018}. \citet{jonsson:2018} evaluates the performance of ASPCAP by comparing the stellar parameters and abundances derived to those of high-resolution optical studies of giants. The typical uncertainties found in \teff, \logg\ and $\rm [Fe/H]$ are in the order of $\rm \sim 100\,K$, $\rm \sim 0.2\,dex$ and $\rm \sim 0.1\,dex$, respectively. Similar uncertainties are found in \citet{schultheis:17} by comparing to other spectroscopic results for bulge giants in Baade’s Window. However, since both these comparisons were made for mainly stars with \teff $>3800\,K$ (\citet{schultheis:17} have two cooler comparison stars), these results cannot strictly be extrapolated to our much cooler M-giants.

Since \teff\ is the parameter having the greatest influence on the derived Sc abundance (see Section \ref{subsec:atomicsctruct} and Figures \ref{scfe_feh_solar}, \ref{scfe_teff_solar}, and \ref{eqwidthvsteff}), we go on evaluating the ASPCAP accuracy of this parameter further. Table~\ref{tabphot} shows a comparison of the effective temperature measured by APOGEE and  the photometric \teff\ based on the $\rm J-K_{S}$ vs. \teff\ relation from \citet{houdashelt2000}. In addition, we queried the Gaia DR2 database and found  Gaia temperature estimates for seven of our objects. The temperatures from Gaia DR2 were determined by using the $\rm G_{BP} -G$  and $\rm G_{RP} - G$ colors as inputs together with a training set of different labels from different spectroscopic surveys such as APOGEE, RAVE, LAMOST, and the  Kepler Input Catalogue. For a more detailed description of the Gaia DR2 temperatures, we  refer to \citet{andrae2018}. The mean difference between the spectroscopic and photometric temperatures from \citet{houdashelt2000} is $\sim 50\,$K with a standard deviation of 240\,K. The mean difference between APOGEE and Gaia is $\sim 90\,$K with a a standard deviation of 185\,K. We can conclude that the spectroscopic temperatures of our cool M giants derived from APOGEE are consistent with those of photometric measurements and Gaia's estimated temperatures, and hence expected not to show any large systematic uncertainties that in turn would skew our abundance analysis.

\begin{table}
\caption{Comparison of effective temperatures between APOGEE, photometric temperature and those of the Gaia DR2 release.}
\label{tabphot}
\begin{tabular}{lcccc}
Star & \teff$^{\textrm{APOGEE}}$ & \teff$^{\textrm{photometric}}$ & \teff$^{\textrm{Gaia}}$ \\
\hline
2M17584888--2351011 & 3652 & 3673 &  --  \\
2M18103303--1626220 & 3436 & 3350 & 3282 \\
2M18191551--1726223 & 3596 & 3191 & 3297 \\
2M18550791+4754062  & 3915 & 3870 & 4000 \\
2M19122965+2753128  & 3263 & 3402 & 3286 \\
2M19411424+4601483  & 3935 & 4238 & 4048 \\
2M21533239+5804499  & 3708 & 3451 & 3423 \\
\end{tabular}
\end{table}

For $\alpha$ Boo the Gaia FGK benchmark stars parameters are used \citep{heiter:14,heiter:15,jofre:15} with uncertainties in \teff, \logg\ and $\rm [Fe/H]$ being $\rm 35\,K$, $\rm 0.09\,dex$ and $\rm 0.08\,dex$, respectively.

The stellar parameters of the three Galactic Center stars have been determined by \citet{rich:17}. The \teff\ is determined from the strength of the CO bandhead in low-resolution, K-band spectra \citep{schultheis:16}. The uncertainties in \teff, \logg\ and $\rm [Fe/H]$ are in the order of $\rm \sim 150\,K$, \mbox{$\rm \sim 0.3\,dex$} and $\rm \sim 0.2\,dex$, respectively. 

\myedit{For determination of microturbulence we look to the abundance investigations of M giants in the solar neighborhood based on high-resolution near-IR spectra \citep{smith:85,smith:89,smith:90}, and for red giants in the Large Magellanic Cloud \citep{smith:2002}. The microturbulence was found by demanding that abundances from lines from a few atomic species be independent of strength, with values from $2-3\,$km\,s$^{-1}$ for M giants. A more recent detailed analysis of spectra of five red giant stars by \citet{smith:13} is used to provide an empirical relation described by \citet{rich:17} for estimating microturbulence for our stars. The microturbulence values used are listed in Table~\ref{tab:star}.} 

\subsection{Scandium Abundance Determination}

To determine the scandium abundance of the analyzed stars we find best fitting synthetic spectra based on radiative transfer and line formation calculated for a given model atmosphere  \citep[MARCS atmosphere models in spherical geometry; ][]{marcs:08} assuming the fundamental stellar parameters shown in Table~\ref{tab:star}. We use the code Spectroscopy Made Easy (SME) for synthesizing model spectra and finding best fit to the observed spectra using $\chi^2$ minimization \citep{sme,sme_code}.

We use four neutral scandium (\ion{Sc}{1}) features seen in the spectra of the analyzed stars for determining scandium abundances. The approximate wavelength in air of these features are 21730~Å, 21812~Å, 21842~Å, and 22394~Å, where the first three are visible in the spectra plotted in Figure~\ref{2mvsgc}. Due to hyperfine splitting of neutral scandium the four spectral features consists of a total of 59 scandium lines, making the analysis robust against saturation effects that could otherwise be suspected due to the strength of the spectral features.

\begin{figure}[t]
\centering
\includegraphics[width=0.5\textwidth]{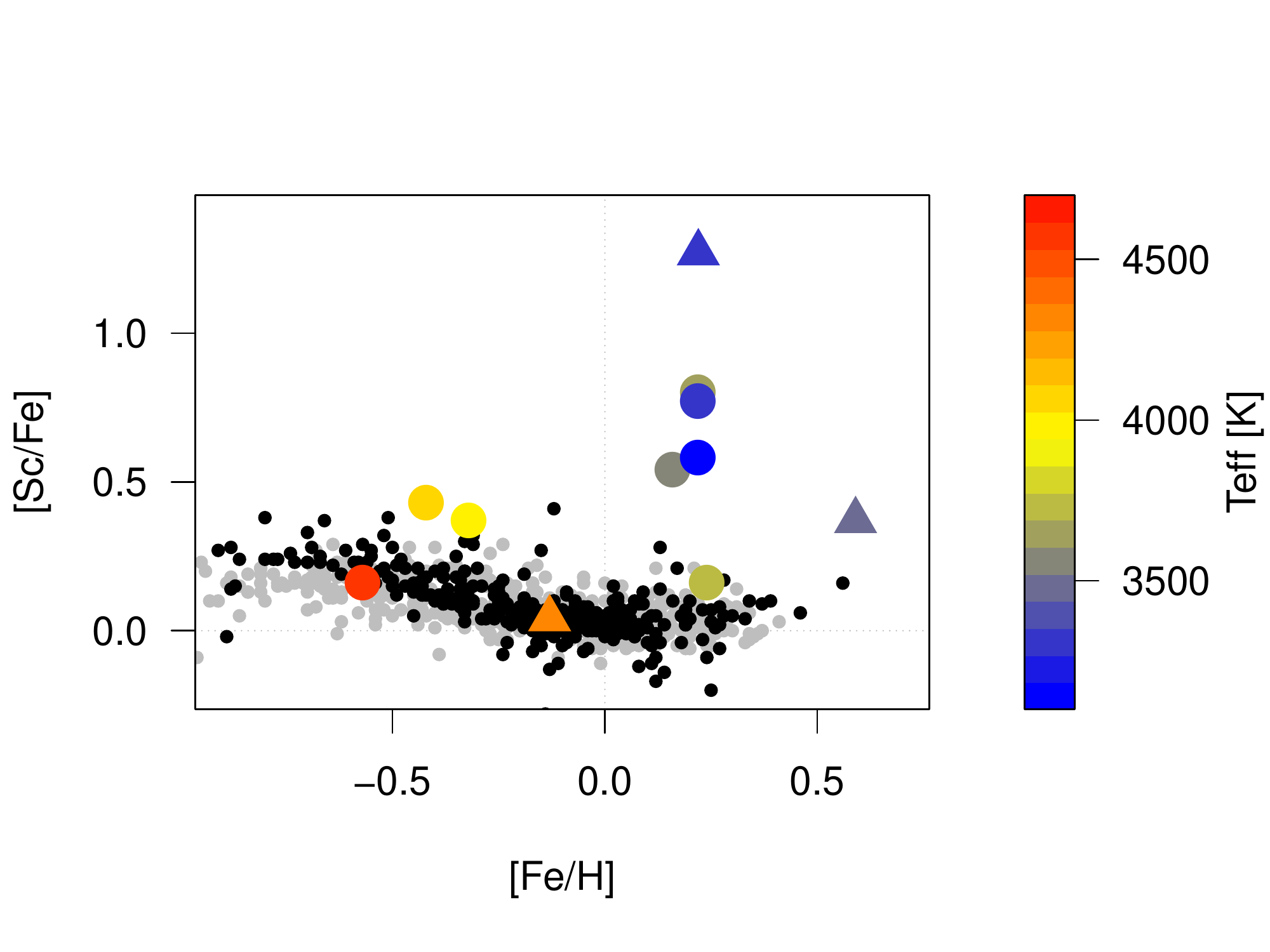}
\caption{$\rm [Sc/Fe]$ vs.\ $\rm [Fe/H]$ for stars in the solar neighborhood (disks) and Galactic Center (triangles), colored by their effective temperature. Shown also are the scandium abundances derived from local disk giants from \cite{lomaeva:18} as black dots and the scandium abundances derived from disk dwarfs by \citet{Battistini15} as gray dots. Note how the derived $\rm [Sc/Fe]$ abundance ratio increases drastically as the stars becomes cooler, while the stars near 4000~K and above are more in agreement with both \cite{lomaeva:18} and \cite{Battistini15}.}
\label{scfe_feh_solar}
\end{figure}

\begin{figure}[t]
\centering
\includegraphics[trim={0cm 1cm 0.5cm 3cm},clip,width=0.5\textwidth]{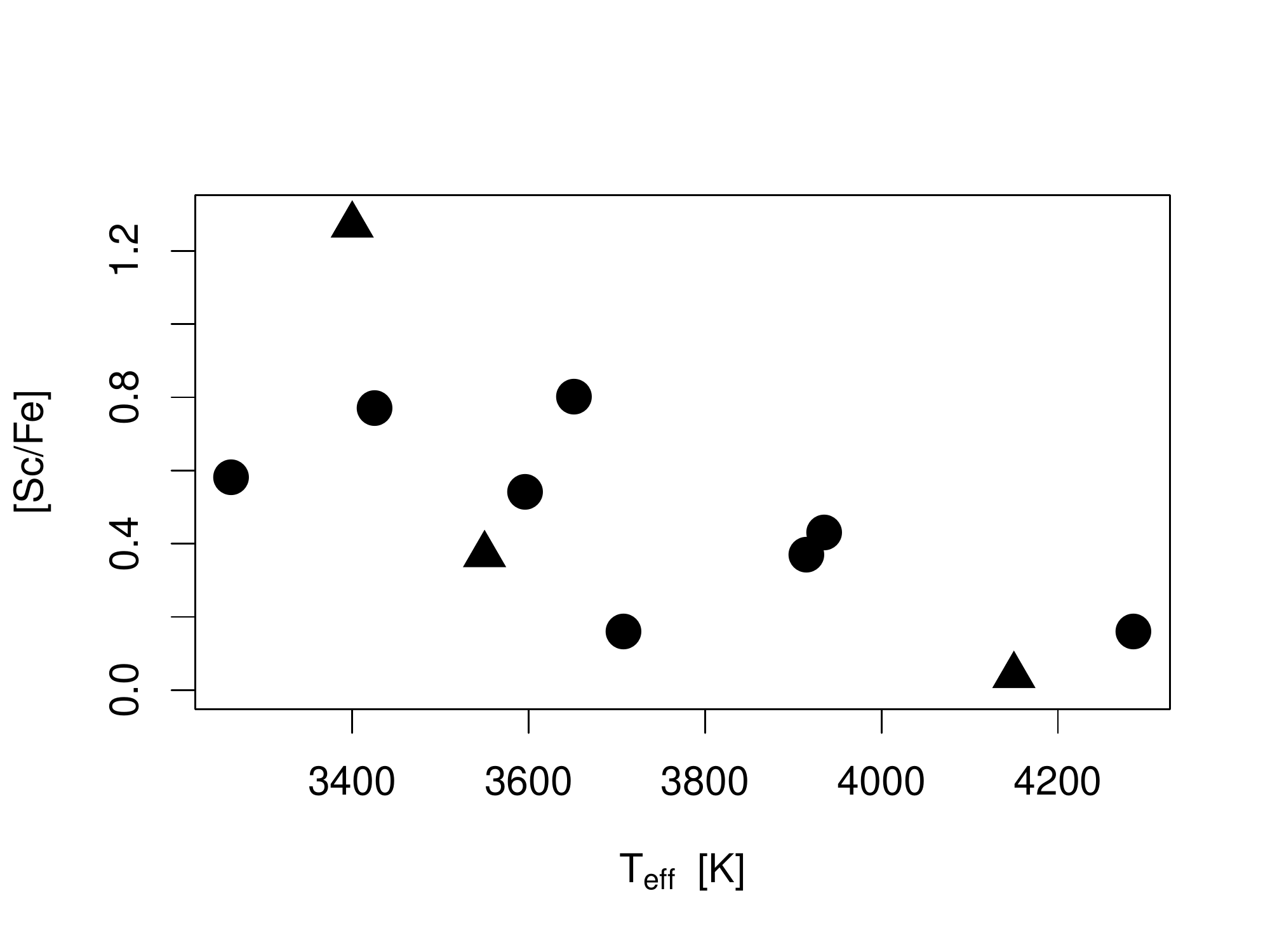}
\caption{[Sc/Fe] vs.\ $T_\textrm{eff}$ for stars in the solar neighborhood (disks) and Galactic Center (triangles). Notice that both populations follow approximately the same trend. This plot shows that the derived scandium abundance simply increases with lower temperatures, suggesting that the derived ``abundances'' are spurious, being systemically in error, perhaps due to an incomplete model of the line formation.}
\label{scfe_teff_solar}
\end{figure}

The determined scandium abundances are shown in Table~\ref{tab:star} and plotted in Figure~\ref{scfe_feh_solar} as a function of [Fe/H] and in Figure~\ref{scfe_teff_solar} as a function of temperature. The uncertainties in [Sc/Fe] are on the order of $\rm \sim 0.1\,dex$, assuming that the model assumptions holds.


\section{Results and Discussion}

\subsection{Observed Spectra}

Our central result is given in Figure~\ref{2mvsgc} which shows an array of our Nuclear Star Cluster K band spectra interleaved with our identically observed local giants of similar spectral type (i.e.\ very similar stellar parameters). It is clear from this figure that the M~giant spectra are overall quite similar, even taking into account the small differences in stellar parameters. This similarity extends to the many of the atomic lines as well. At first glance, the plot suggests that the scandium abundance simply increases with lower temperature; this begins to point toward an explanation other than a true enhancement in scandium abundance, such as being unable to accurately model the line formation. Similar enhancements for vanadium and yttrium are also clear; both of these lines strengthen dramatically below 3600 K. We can conclude based only on empirical grounds that there is room for only very small differences in the abundances of these elements between stars in the Nuclear Cluster Stars and stars in the solar neighborhood. In particular, the scandium abundance is very unlikely to be anomalously elevated in the Galactic Center Nuclear Star Cluster, as asserted in \citet{do:18}. 

\subsection{Derived Scandium Abundances}

Figure \ref{scfe_feh_solar} shows our derived scandium abundances in the form of the $\rm [Sc/Fe]$ trend versus metallicity, i.e.\ the $\rm [Fe/H]$ abundances. We show our seven M~giants in the solar neighborhood together with $\rm \alpha$-Boo and three M~giants in the Nuclear Star Cluster \citep[from][]{rich:17}, colored as a function of effective temperature. As a background trend, we also plot the $\rm [Sc/Fe]$ abundance trends found for the solar neighborhood, determined from optical \ion{Sc}{2} lines from \cite{joensson:17b,lomaeva:18} for K~giant spectra and from \citet{Battistini15} for dwarf star spectra; these trends follow approximately that of a typical $\alpha$-like element in the solar neighborhood, with elevated $\rm [Sc/Fe]$ at $\rm [Fe/H]<-0.5$ and a gentle decline toward Solar $\rm [Sc/Fe]$  It is evident that these well-established trends are very different from our derived $\rm [Sc/Fe]$ values based on \ion{Sc}{1} lines in our M giant spectra.  Our derived abundance ratio determined for $\rm \alpha$-Boo do agree with the general trend.  It is evident that two giants with the highest temperature (from both the NSC and Local giants) fall exactly on the general trend even though their abundances are 0.5 dex apart.  All of the cooler stars fall above the trend, and it is evident that the coolest stars are 0.5 to 1.3 dex above the trend, with the coolest stars showing the most extreme derived $\rm [Sc/Fe]$ ``abundance'' values.

Indeed, the most striking features seen in Figure \ref{scfe_teff_solar} is how our derived $\rm [Sc/Fe]$ ``abundance'' ratios increase dramatically with decreasing 
\teff\ leading to anomalously high scandium abundances of up to 1.3 dex for the cooler stars (\teff$ < 3800\,$K), while the warmer stars show more normal $\rm [Sc/Fe]$. There is no astrophysical basis for the scandium abundance to depend on the {\it temperatures} of the stars, and therefore we can safely conclude that the derived abundances for the cooler stars are either plagued with very large systematic uncertainties or based on assumptions in the abundance determination that are not valid. This calls for a discussion of the physical processes involved.

\myedit{
We note that \citet{smith:2002} did not derive any anomalously high scandium abundances from their high-resolution, near-IR spectra of 12 M giants in the Large Magellanic Cloud (LMC) with effective temperatures in the range of 3600--4000\,K and metallicities between --1.1 and -0.3\,dex. The \ion{Sc}{1} line they used at 23404.8\,Å, which is not included in our wavelength range, is measured to have equivalent widths in the range of 70--650\,mÅ. This is comparable to the equivalent widths of similar lines we find for our stars in the same temperature range. Apart from scandium, \citet{smith:2002} also derived the abundances of sodium, titanium, and iron, all from neutral lines. The [Na/Fe] and [Ti/Fe] abundance trends decrease with iron slightly below the disk trends which is to be expected for the LMC. However, their [Sc/Fe] trend does not show such a depletion, possibly indicating higher-than-expected scandium abundances derived. Finally, \citet{smith:2002} finds that the scandium abundance determination is very temperature sensitive with an uncertainty of almost 0.2\,dex for a shift in 100\,K in the effective temperature, which agrees with the modeled temperature sensitivity which we find to be above 0.1\,dex, as discussed earlier in Section~\ref{teffeffect}.

 
}


\subsection{Understanding the Strong Scandium Lines}

As already noted in their discussion of spectral lines in M giants in the Nuclear Star Cluster \citep{rich:17}, the \ion{Sc}{1} lines observed in the K-band are indeed stronger than expected from an LTE spectral synthesis analysis for a reasonable scandium abundance. In this context, it is therefore of interest to consider possible processes and properties of these scandium lines that might be responsible for these anomalously high line strengths.

In LTE and for unblended, weak lines, i.e.\ lines that are not saturated, the observed line strengths are directly proportional to the atomic line strengths (Einstein $A$ coefficients or $\log gf$-values) and the number density abundance ($A_\mathrm{element}=\log N_\mathrm{element}$[cm$^{-3}$]) of the element causing the spectral line. In order to derive the abundance it is therefore vital with  a well determined $\log gf$-value. We therefore use the recent laboratory measurements of \citet{pehlivan:sc} for all our lines. This ensures a minimum uncertainty due to the intrinsic line strengths of the \ion{Sc}{1} lines.

The strength of saturated lines can, on the other hand, also be very sensitive to the assumed microturbulence \citep[see e.g.][]{gray:2005}, leading to large uncertainties in the derived abundances. However, having an odd nuclear spin, the scandium lines are subdivided into many components due to the hyperfine structure (HFS). In the case of our \ion{Sc}{1} lines, the HFS has the effect of delaying the onset of saturation (since the strong line feature consists of many weak lines), but also to change the appearance of line. Further, the pressure broadening effect should be smaller on the components compared to an equally strong single line. We have, therefore, taken into account the HFS in our synthesis, reproducing successfully the line profile. Furthermore, our modeled lines are indeed relatively insensitive to microturbulence, even though the components have a relatively large summed equivalent width.

A further concern when working with spectral lines that yield anomalously high abundances, but also with lines in metal-rich and especially cool stars in general, is the contribution of unaccounted blends from molecular and atomic species. The accurate wavelength of molecular lines blending is a concern, but if the wavelength are known precisely enough, these lines may be properly accounted for. The CN line list we use \citep{sneden:14} is very precise, so such blends can be accurately taken into account.  Blended atomic lines might make an abundance analysis impossible; the existence of such blends might be known, but their atomic data are not accurately known, or there might be unknown blending lines which are not taken into account at all. However, in our spectra all four \ion{Sc}{1} lines are stronger than modeled by approximately the same amount. Therefore, it is very unlikely that all the four lines would be blended at the same time. We therefore conclude that the \ion{Sc}{1} lines are not significantly affected by blends.

\ion{Sc}{1} lines are known to be temperature sensitive, and one concern would be that the anomalously strong \ion{Sc}{1} lines are caused by a systematically incorrect temperature scale. To assess this possibility, we have determined the $\rm [Sc/Fe]$ abundances and the equivalent widths from a grid of stellar models. We have focused on the temperature range of M giants ($3000<$\teff /[K]$<4000$) since these are optimal objects to be observed in the Nuclear Star Cluster. A surface gravity is determined for a given (\teff, \logg) combination from the Yale-Yonsei (YY) isochrones \citep{yy}.  Figure \ref{scfevsteff} shows the derived Sc number-density abundances from a line of a typical equivalent width of 300\,m\AA. A decrease of 100\,K in \teff\ increases the derived Sc abundances by more than 0.1\,dex. The scandium lines are indeed temperature sensitive, but this alone cannot explain the systematically stronger lines for our cooler stars. There is no reason to believe that the cooler solar neighborhood stars have \teff\  systematically too high by up to 800K.

We conclude that our derived abundances, based on an {\it LTE} abundance analysis, are precise and accurate to within at least 0.2\,dex, with the temperature sensitivity being the largest source of uncertainty. However, we can also conclude that stars below approximately \teff$<3800$\,K should not be used for deriving stellar scandium abundances from these lines based on a traditional LTE analysis.  We have demonstrated the strong temperature dependence but we cannot specify which model assumptions are invalidated, complicating any abundance analysis.  However, with its low ionization potential, \ion{Sc}{1} is a minority ionization stage throughout the photosphere, and any departures from LTE (and 3D effects) will affect \ion{Sc}{1} lines more than ionized lines \citep{asplund:05}. Changes in the ionization rates, and therefore the ionization balance, change the relative population of the minority species more that that of the majority species. 

No non-LTE investigations of scandium have been made for cool giants (to our knowledge), but \citet{zhang:08} investigated the formation of {\it optical} scandium lines in the Sun and found large departures from LTE for \ion{Sc}{1} lines and none for ionized lines. The \ion{Sc}{1} lines in the Sun are weaker than expected from LTE, whereas we see the opposite for M giants. Whereas for metal-poor, cool stars, photoionization is a major process that can under-populate neutral atoms \citep{gehren:01}, leading to weaker lines in non-LTE, for collision-dominated ions \citep{gehren:01}, photon suction and over-recombination tend to strengthen the lines instead. An over-recombination can be caused by the smaller mean-intensity-field compared to the Planck function, which is the case in the near-IR \citep{asplund:05}. These are processes that could cause stronger lines than expected from an LTE treatment. 

\myedit{The treatment of convection when using a full 3D hydrodynamical modeling instead of using a traditional 1D approach could also significantly affect the line formation. For neutral lines in the near-IR formed in red giants discussed in \citet{kucinskas:13,cerniauskas:17}, which are slightly warmer than ours, the abundance corrections are shown, however, to be small.}

A strong suspicion is thus that \ion{Sc}{1} lines formed in cool M giant atmospheres also are affected by non-LTE effects. In their non-LTE study of titanium lines, with a similarly low ionization potential as for scandium, \citet{bergemann:12} show that the J-band lines in supergiants are indeed strengthened in non-LTE. The effect increases for increasing metallicities and decreasing temperatures, from a negligible effect at 4200\,K to 0.4\,dex at 3400\,K, their lowest temperature point. Above 4200\,K the opposite effect is in play. If the same processes are important for scandium, these trends could explain also the strengthening of the scandium lines for cool, metal-rich stars. A detailed non-LTE study is, however, required to find the dominant process in play for the scandium lines in the K-band for M~giants. However, in general, it is difficult to isolate the dominant process causing a departure of a Boltzmann-Saha level populations for a transition in a complex atomic structure \citep{bergemann:12}.

\subsection{Vanadium and Yttrium Lines in the K-Band}

In the K-band, there are in addition a number of very strong vanadium and yttrium lines, see for example the spectra plotted in Figure~\ref{2mvsgc}. \citet{do:18} suggest these elements to be significantly enhanced at the Galactic Center, also with a potentially very different chemical enrichment history  there. Based on the similarity of the solar neighborhood stars we present here, one might suspect that the \ion{V}{1} and \ion{Y}{1} lines might suffer from the same line strengthening effects as the scandium lines do.

Yttrium is homologous to scandium, residing just below scandium in the periodic table. It has the same atomic structure but with principle quantum numbers one unit larger and thus 4d and 5s instead of 3d and 4s for \ion{Sc}{1}, which makes \ion{Y}{1} lines of the configurations 4d$^2$5s--4d5s5p appear at low excitation energies. The strength of the \ion{Y}{1} line should therefore react in the same way to the line formation process, and is thus expected to be stronger in the same way as in the case of \ion{Sc}{1}. In addition, lanthanum residing in the period below yttrium, would show a similar effect for the lines 5d$^2$6s--5d6s6p, had there been lines detected in the K-band. Only the relative abundance differs. Laboratory studies of \ion{Y}{1} and \ion{La}{1} are ongoing.

Vanadium is the next odd element after scandium, lying after titanium in the periodic table. Its structure more closely resembles that of the iron group elements, such as \ion{Fe}{1} with the majority of the strong transitions appearing at high excitation. There are exceptions, such as the transition at 3d$^5$~a$^6$S--3d$^5$($^5$D)4p z$^6$P at 22493\,Å. A non-LTE strengthening of vanadium cannot be excluded. We infer that non-LTE strengthening is the best option to explain the odd behavior of vanadium, given the arguments advanced here for scandium.

\section{Conclusions}

Our question in this paper is whether the abundance of scandium  in stars observed in the Nuclear Star Cluster is really unusual. An unusual abundance would have a profound effect on our understanding of the formation of the Nuclear Star Cluster and its stellar population.
There are several physical reasons for a spectral line to be strong, and by going through all these, we conclude that the strong \ion{Sc}{1} lines in the Nuclear Star Cluster most likely are due to line formation effects and certainly not due to an anomalous high scandium abundance. These \ion{Sc}{1} lines in the K-band are not good abundance diagnostics for the elements in cool M giants. Our conclusion is based on the fact that similar stars in the solar neighborhood, where the scandium abundances are known, show similarly strong \ion{Sc}{1} lines, much stronger than they can be modeled with a traditional synthetic spectroscopy. We conclude that the \ion{Sc}{1} lines in cool stars are strong everywhere in the Galaxy and is an inherent property of the line formation process, and should not be used to derive the scandium abundance of the stars. Lines of ionized scandium should be used instead or studies should use warmer stars. Non-LTE calculations for scandium, and perhaps other physical phenomena needing theoretical modeling, are needed before we can use the \ion{Sc}{1} lines at all.

Our findings emphasize the perils of attempting to infer composition from a small number of lines that are far too strong for any conventional application of abundance analysis. Although high resolution infrared spectroscopy makes it possible to study the composition of stars in highly obscured regions, the relative paucity of weak lines requires that work in the infrared is approached with great caution. In the case where anomalies are suspected, it is important to turn to general established trends derived from optical measurements in the disk and bulge, as an additional essential check before claiming to discover unusual composition. We also note that at high metallicities, it would be quite surprising to see large enhancements of any species, that would imply extreme productions of individual elements in a metal rich environment. 

The cool giants we have analyzed are almost certainly on the asymptotic giant branch (AGB) and as such, they are likely to be the most luminous stars in their stellar population, contributing a substantial fraction of the K band integrated light.  We know that the NSC is a complex stellar population with a wide range of age and we might expect many extragalactic nuclear star clusters to also have wide age ranges and possibly host very substantial populations of cool, luminous AGB stars.  Individual supergiants can also contribute a substantial portion of the integrated light in the K band.  There is a growing trend to attempt to infer detailed abundances (e.g. abundances of individual atomic species) from the integrated light of stellar populations.  The temptation is strong for the extragalactic nuclear star clusters of low luminosity spirals, which have relatively low broadening due to the stellar velocity dispersion, making them very attractive targets for spectrum synthesis.  We emphasize that if such a practice is carried out for the integrated light in the K band, that it be done using actual spectra of Solar neighborhood giants and a full population synthesis code.   The numbers of cool AGB stars are small, and their contribution to the stellar population is stochastic \citep{frogel:90}.
Even then, our understanding of line formation for cool giants is far from complete and therefore, we suggest that integrated light studies using only the K band be avoided.  

High resolution spectroscopy in the near-infrared is a relatively new subject area with great promise.  The advent of very high resolution spectrographs on the current generation of 8-10m telescopes and future plans for ELTs give the subject great promise.  It is important to support this endeavor with an equally serious investment in laboratory measurements, so that full advantage can be taken of the expected bounty of high quality data.

\acknowledgments
B.T. and N.R. acknowledges support from the Swedish Research Council, VR (project number 621-2014-5640), Funds from Kungl. Fysiografiska Sällskapet i Lund. (Stiftelsen Walter Gyllenbergs fond and Märta och Erik Holmbergs donation), and from the project grant “The New Milky” from the Knut and Alice Wallenberg foundation. M.S. acknowledges the Programme National de Cosmologie et Galaxies (PNCG) of CNRS/INSU, France, for financial support. H.J. acknowledges support from the Crafoord Foundation, Stiftelsen Olle Engkvist Byggm\"astare, and Ruth och Nils-Erik Stenb\"acks stiftelse. R.M.R. acknowledges support from the U.S. National Science Foundation, from grants AST-1413755 and AST-1518271.
This work has made use of data from the European Space Agency (ESA) mission {\it Gaia} (\url{https://www.cosmos.esa.int/gaia}), processed by the {\it Gaia} Data Processing and Analysis Consortium (DPAC, \url{https://www.cosmos.esa.int/web/gaia/dpac/consortium}). Funding for the DPAC has been provided by national institutions, in particular the institutions participating in the {\it Gaia} Multilateral Agreement.

The data presented herein were obtained at the W. M. Keck Observatory, which is operated as a scientific partnership among the California Institute of Technology, the University of California and the National Aeronautics and Space Administration. The Observatory was made possible by the generous financial support of the W. M. Keck Foundation. The authors wish to recognize and acknowledge the very significant cultural role and reverence that the summit of Mauna Kea has always had within the indigenous Hawaiian community.  We are most fortunate to have the opportunity to conduct observations from this mountain.

\facilities{KECK:II (NIRSPEC)}

\software{SME \citep{sme,sme_code}, BSYN \& EQWIDTH \citep{marcs:08}}


\bibliographystyle{yahapj}
\bibliography{references}

\end{document}